\documentclass[english,aps,prb,floatfix,twocolumn,showpacs]{revtex4}
\usepackage[T1]{fontenc}
\usepackage[latin1]{inputenc}
\usepackage{amsmath}
\usepackage{babel}
\usepackage{graphicx}
\usepackage{amssymb}
\usepackage{multirow}

\begin{document}
\title{Origin of Magnetism and trend in $T_{c}$ in Cr-based double perovskites: Interplay of two driving mechanisms}
\author{Hena Das$^{1}$, Prabuddha Sanyal$^{1}$, T. Saha-Dasgupta$^{1,\ast}$,  D.D. Sarma$^{2, \dagger}$}
\affiliation{$^{1}$S.N. Bose National Centre for Basic Sciences,
Kolkata 700098, India}
\affiliation{$^2$ Solid State and Structural Chemistry Unit,
Indian Institute of Science, Bangalore-560012, India}
\date{\today}

\begin{abstract}
Employing first principles density functional calculations, together with solution of the 
low-energy, model Hamiltonian constructed in a first principles manner, we explored the origin of magnetism 
and T$_c$ trend in Cr-based double perovskite series, Sr$_2$CrB$'$O$_6$ (B$'$=W/Re/Os). Our study shows that the
apparently puzzling $T_c$ trend in  Sr$_2$CrB$'$O$_6$ (B$'$=W/Re/Os) series 
can be understood in terms of the interplay of
the hybridization driven mechanism and the super-exchange mechanism. 
\end{abstract}
\maketitle
%%%%%%%%%%%%%%%%%%%%%%%%%%%%%%%%%%%%

\section{Introduction}

 Oxides with high magnetic transition temperature ($T_{c}$) are important for technological advancement. 
A much discussed family of compounds in this connection are double perovskites, 
with a general formula A$_{2}$BB$'$O$_{6}$, A=alkaline/rare-earth metals and B/B$'$=transition metals. 
The observation\cite{sfmo-nature} of  large magnetoresistance with a fairly high ferromegnetic (FM) $T_c$ of about 410 K in 
Sr$_{2}$FeMoO$_{6}$ (SFMO) and its unusual origin of magnetism  brought this family of compounds into the forefront 
of activity. Since then the question has been, 
can $T_{c}$ be boosted even further and if so, is there a systematic trend that
can be observed and understood. Attempts have been made to dope SFMO with La to increase the $T_c$.  Though some partial
success has been achieved\cite{navarro} following this path, $T_c$ was found to be boosted much more efficiently by moving to 
different choices of B and B$'$ ions. The microscopic understanding of this increase, however, has not been achieved. 
A collection of measured transition temperatures seems to bear some correlation with the number of valence electrons among 
different double perovskite compounds (see Fig.1 in Ref.\onlinecite{claudia}). Focusing onto Cr-based series, namely 
Sr$_{2}$CrB$'$O$_{6}$ (B$'$ = W/Re/Os), the family with spectacularly high $T_c$, the measured $T_c$ shows a rapid increase
as one moves from Sr$_2$CrWO$_6$ (SCWO)\cite{tc-w} with $T_c$ $\approx$ 450 K to Sr$_2$CrReO$_6$ (SCRO)\cite{tc-re} with 
$T_c$ $\approx$ 620 K  to Sr$_2$CrOsO$_6$ (SCOO)\cite{tc-os} with $T_c$ $\approx$ 725 K. 
This rapid increase is curious considering the hybridization driven (HD) mechanism\cite{sfmo-prl,terakura} of magnetism accepted so far 
for double perovskites, proposed in the context of SFMO, which would predict decrease rather than the increase.\cite{millis}.
Furthermore, the compound with the highest $T_c$ in this series, SCOO, is found to be an insulator making the issue even more 
interesting. Also, a distinction between others like SFMO and La doped SFMO and this series is a possible role of spin-orbit 
coupling (SOC) due to the presence of 5d elements.

We have used density functional theory (DFT) based calculations together with exact diagonalization of Cr-B$'$ model Hamiltonian 
constructed in a first-principles-derived Wannier function basis, in order to probe this issue. Though few DFT based studies\cite{dft-study} 
exist for individuals of these compounds, to best of our knowledge, theoretically no comprehensive study exists to address the origin of 
growing trend in $T_c$ within the series.

\section{Methodology}

 The DFT calculations were carried out using the plane wave pseudopotential method implemented 
within Vienna Ab-initio Simulation 
Package(VASP).\cite{vasp} The exchange-correlation functionals were approximated by generalized gradient approximation (GGA).\cite{GGA} We have used projected augmented wave (PAW) potentials \cite{PAW} and the kinetic energy cut-off for expansion of 
wavefunctions used was 450 eV. Reciprocal space integrations have been carried out with a k-space mesh of 6$\times$6$\times$6.
For extraction of a few-band, tight-binding Hamiltonian out of full GGA calculation, we have carried out muffin-tin orbital (MTO)
based NMTO-downfolding calculations.\cite{NMTO} The reliability of the calculations in two basis sets has been cross-checked. 

\section{Results}

\subsection{Examination of Basic Electronic Structure}

\begin{figure}
\includegraphics[width=8cm]{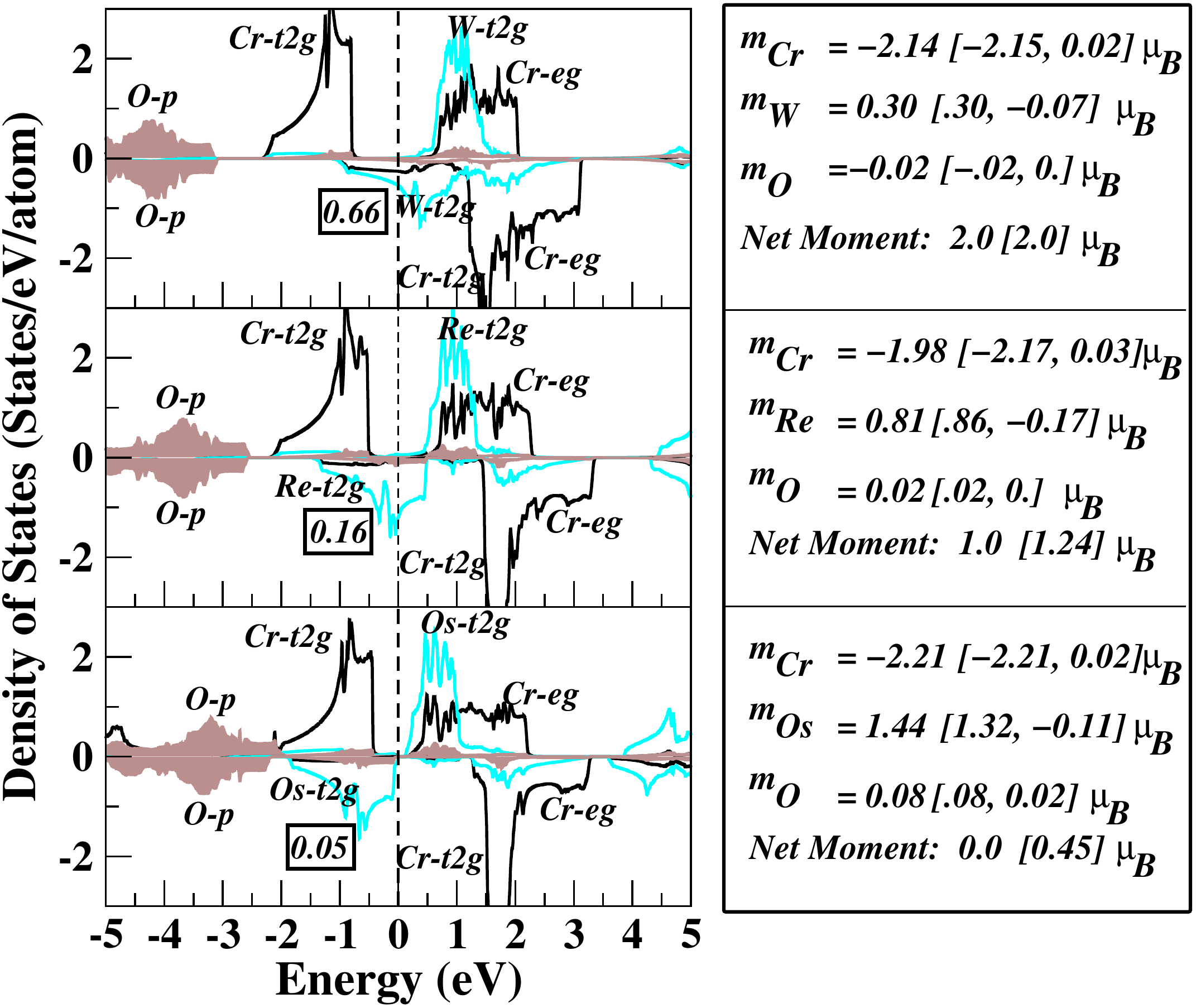}
\caption{\label{fig-1} (Color online) Left Panels: GGA DOS projected onto Cr $d$ (black solid lines), 
B$'$ $d$ (Red/Grey solid lines) and O $p$ (shaded area). Zero of the energy is set at $E_{F}$.  The
numbers within the boxes indicate the Cr t$_{2g}$ contribution in the bands crossing $E_F$, with respect 
to that of B$'$. Right Panel: Calculated net magnetic moment and 
magnetic moments at Cr, B$'$ and O sites. The numbers within the bracket denote the result of GGA+SO calculations,
the first entry being the spin moment and the second entry being the orbital moment. From top to bottom, the plots 
correspond to SCWO, SCRO and SCOO respectively.}
\end{figure}  

In order to unravel the origin of magnetism in Sr$_2$CrB$'$O$_6$ series, let us first critically examine 
the electronic density of states (DOS) of these compounds. Left panel of Fig.1 shows the DOS, as obtained in 
spin-polarized DFT calculations within GGA.
The states close to Fermi level ($E_{F}$) are dominated by Cr and B$'$ $d$ states hybridized with 
O $p$ states, while the O $p$ dominated states separated from Cr and B$'$ $d$ dominated states occupy the energy range
 far below $E_{F}$ and Sr $s$ and $d$ dominated state remain far above $E_{F}$.  
The $d$ states of Cr and the B$'$ ions are exchange split as well as crystal field split. 
The empty B$'$ $t_{2g}$ states in the up-spin channel appear in 
 between the crystal field split Cr $t_{2g}$ and $e_{g}$ states, gaped from $E_{F}$
 while the B$'$ states in the down spin channel  hybridized with
 Cr $t_{2g}$ states, either cross the Fermi level as in the case of W and Re compounds, or remain completely occupied, 
as in case of the Os compound. It is rather intriguing to notice that the hybridization between Cr $t_{2g}$ and B$'$ $t_{2g}$
in the down spin channel, progressively gets weakened in moving from W to Re to Os compound. This may be appreciated
by considering the Cr contribution measured with respect to B$'$ contribution
for the states in the down spin channel close to $E_{F}$. For W, Re and Os compounds, it is found to
be 66$\%$, 16 $\%$ and 5$\%$ respectively. This is caused by the gradual moving down of the B $'$ energy level, in moving from left
to right of the periodic table across the same row (W $\rightarrow$ Re $\rightarrow$ Os), reflecting an increase in the ionic
potential experienced by the 5d electrons with an increase in nuclear charge. As discussed later, the hopping interaction connecting Cr and B$'$ $t_{2g}$ states on the other hand remains similar across the series.
Analysing the DFT
calculated magnetic moments, presented in the right panel of Fig.1, we find 
the second interesting observation that the magnetic moment per electron at B$'$ site, 
defined as $m/d$, where $m$ is the calculated moment and $d$ is the valence
count at B $'$,\cite{footnote1} keeps growing from W to Re to Os.
The increase in $m/d$ becomes even more evident taking into account the magnetic moment at O site,
which is small and point to Cr moment for SCWO, small (large) and point to Re (Os) moment for SCRO (SCOO).
This prompts us to conclude that there is a growing intrinsic moment
that develops at B $'$ site following the dehybridization effect between Cr and B$'$.

\subsection{NMTO-downfolding calculations}

\begin{figure}
\rotatebox{0}{\includegraphics[width=7.6cm]{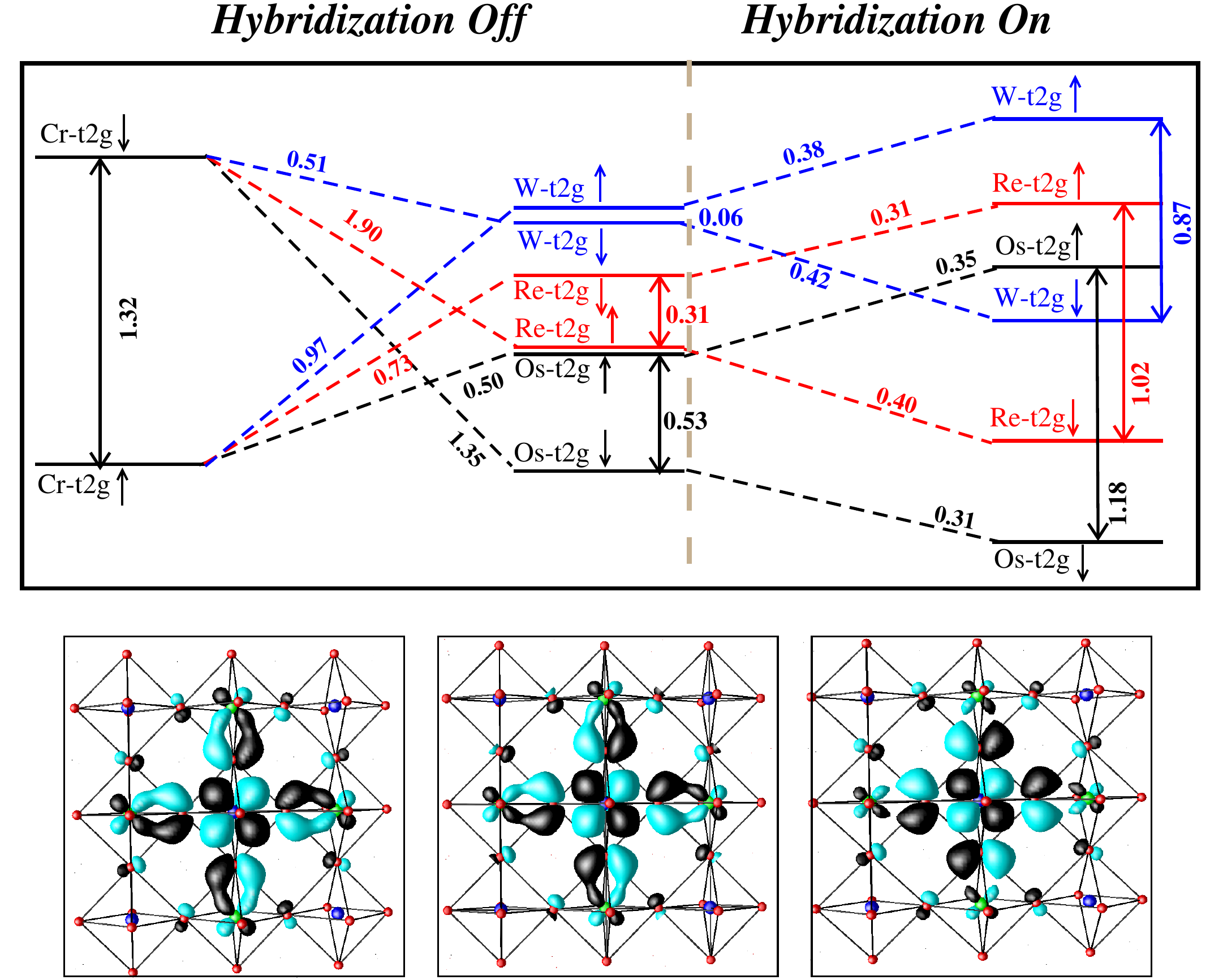}}
\caption{\label{fig-2} (Color online) The energy level diagram (upper panel) 
and massively downfolded Wannier functions (lower panels)
for Sr$_{2}$CrB$'$O$_{6}$ series. For Wannier function plots, constant value surfaces have been plotted with two oppositely signed lobes colored differently. From left to right in lower panel, the plots correspond to SCWO, SCRO and SCOO respectively. The numbers in the energy level diagram are in unit of eV.}
\end{figure}  

In order to further analyse the findings of the electronic structure calculations, we have carried out NMTO downfolding calculations, which is engineered to define energy-selected, effective 
Wannier functions by integrating out degrees of freedom that are not-of-interest ({\it downfolding}).
As a first step we downfolded O $p$, Sr as well as Cr and B$'$ $e_g$ degrees of freedom. This defines an effective basis consisting 
of Cr $t_{2g}$ and B$'$ $t_{2g}$ states.
 In the second step, we applied massive
  downfolding, keeping only B$'$ $t_{2g}$ degrees of freedom active and downfolding all
  the rest including Cr $t_{2g}$ degrees of freedom.
On site matrix 
 elements of the real space Hamiltonian defined in the Cr $t_{2g}$-B$'$ $t_{2g}$ basis and the 
massively downfolded basis give the energy level positions before and after switching on the
hybridization between Cr and B $'$ states, respectively. Fig.2 summarizes the results 
for W, Re and Os compounds. The energy levels in the left and right half of the upper panel, demarked with vertical grey line, 
depict the energy level positions obtained out of downfolded Cr $t_{2g}$-B$'$ $t_{2g}$ basis and the 
massively downfolded basis, respectively. 
 The lower panels exhibit the plots of the one of the $t_{2g}$ ($xy$) Wannier functions corresponding to massively
 downfolded Hamiltonian in the down spin channel. Examination of Fig.2 brings out two aspects: Firstly, the progressive 
dehybridization effect, as discussed in the context of DOS plots, is 
evident in the plots of Wannier functions. The central parts of the Wannier functions are shaped according to B$'$ $xy$ 
symmetry and the tails of the Wannier functions sitting at neighboring sites are shaped according to O $p$
 and Cr $t_{2g}$ symmetry. The tails reflecting the hybridization between the Cr $t_{2g}$ and 
 B$'$ $t_{2g}$-O weaken as one moves from W to Re to Os compound. As a consequence, the ratio of renormalized spin splitting to that of the bare
splitting at B$'$ site reduces drastically from W (14.5) to Re (3.3) to Os (2.2). Secondly, considering the level splitting at B$'$ site 
before switching on the hybridization, we find while the splitting at W is negligibly small confirming the nonmagnetic character of the B$'$ site, those for Re and Os are found to be $\approx$ 0.31 eV and 
$\approx$ 0.53 eV, respectively. These values are significantly larger compared to what one would have expected considering
 $d^2$ valence in Re and $d^3$ valence in Os compared to $d^1$ valence in case of W with 0.06 eV splitting, which would have given rise to splittings of 0.12 eV and 0.18 eV 
respectively. This confirms the presence of a growing
intrinsic, local moment at B$'$ site as one moves from W to Re to Os, driven by the dehybridization effect.
The magnetism in Cr-B$^{\prime}$ series, therefore, needs to be understood as an interplay of two mechanisms: HD mechanism as 
operative in SFMO which causes renormalized,
negative spin splitting within B$^{\prime}$ states that appear in between the exchange split Cr $t_{2g}$ states, and the 
superexchange (SE) between the moment at Cr site and the intrinsic moment at B$^{\prime}$ site, which would
align the moments at Cr and B$^{\prime}$ sites antiparallely. For W, the intrinsic moment being negligible, the magnetism
is entirely driven by HD mechanism, while for the other extreme of Os, the hybridization effect
is weak, SE having a rather large contribution. The presence of such intrinsic moment at 5d site is counterintuitive at a first
glance. Comparing the situation, with the double perovskite Sr$_{2}$ScReO$_6$, for which the magnetism has been recently investigated,\cite{njp} Re was found to possess a rather small intrinsic moment of size 0.013 $\mu_B$. Our electronic structure calculations carried out for Sr$_{2}$ScReO$_6$ find also a similarly small moment (0.03 $\mu_B$). The unusual localized aspect of Re or Os, 
in case of Cr based compounds therefore arises due to the relative positioning
of the Cr and B$'$ energy levels, which narrows down the width of the B$'$ states substantially in the up spin channel. The development of the intrinsic
moment is thus helped by the delicate energy level structure responsible for the HD mechanism and would
not have been present otherwise.

\subsection{Total Energy Calculations}

The calculations discussed so far, do not include SOC, which may be important.  The numbers within the bracket in right
panel of Fig.1 show the individual spin and orbital moments as well as net moments as obtained in GGA+SO calculations.
As expected, the orbital moments are large at B$'$ site. Interestingly, we find while the net moment was zero for
SCOO without SOC, it is the consideration of SOC that gives rise to a non-zero moment, due to the uncompensated orbital
moment at B$'$ site. Whether SOC has any influence on the trend within the $T_c$'s, therefore needs to be explored.
Interplay of two driving mechanisms in W-Re-Os series, however, makes it difficult to extract the magnetic exchanges as energy
difference between two specific magnetic configurations. The presence of finite, intrinsic
moment at B$'$ site, as is the case for Re and Os compounds, makes the moment at B$'$ site frustrated in an
antiferromagnetic (AFM) configuration of Cr spins. The finite presence of HD mechanism, on the other
hand, disfavors stabilization of magnetic configurations with majority of B spins aligned parallely to that of the moment 
at B$'$ site. Admitting these difficulties,  we carried out total energy calculations of the two possible spin configurations, one FM 
arrangement and another A-type AFM arrangement of Cr spins, with Cr spins between two adjacent planes are antiferromagnetically coupled, and are ferromagnetically
coupled in plane. For the AFM calculations, the moment at B$'$ site was found to be antiparallely (parallely) 
aligned to the spins of in-plane (out-of-plane) Cr sites which are four (two) in number. 
As expected, the energy differences between FM and AFM configurations are found to be positive for all cases 
proving FM arrangement of Cr spins to be the stable phase. The values of the energy difference is found to increase from W to Re system 
(from 0.23 eV/formula unit to 0.25 eV/formula unit), and then decrease from Re to Os system (from 0.25 eV/formula unit to 0.24 eV/formula unit). 
Introduction of SO interaction though changes the individual energy differences by about 0.03 eV, the trend remains unaltered. This indicates that although the 
presence of substantial SOC at B$'$ site is important for producing the large magneto-optical signals,\cite{moke} it plays little role in setting up the trend in $T_{c}$. In order to examine the role of missing correlation effect in GGA, we have also carried out LDA+U calculations with a choice of U value of 3 eV at Cr site and 0.8 eV at B$'$ site. Application of larger U at Cr site and relatively smaller one at B$'$ sites, as expected from the relative band-widths, were  found to preserve the general conclusions intact with more localized character of d states at B$'$ site, as expected.

\subsection{Exact Diagonalization Study of Model Hamiltonian}

In view of the difficulty in stabilizing the appropriate excited state magnetic configuration, the stability of the 
FM arrangements of Cr spins may be measured as the energy difference between the FM spin configuration and the paramagnetic (PM) phase.
The description of the PM phase needs consideration of different disordered spin configurations and averaging over a large number
of them, which is almost impossible within the DFT framework. Such calculations are much easier to handle within a model
Hamiltonian description.  The model Hamiltonian, describing the interplay of the HD and SE mechanism
 may be written as,
\begin{eqnarray*}
H & = & \epsilon_{Cr}\sum_{i\in B}f_{i\sigma\alpha}^{\dagger}f_{i\sigma\alpha}+
\epsilon_{B'}\sum_{i\in B'}m_{i\sigma\alpha}^{\dagger}m_{i\sigma\alpha} \\
& & -t_{CB'}\sum_{<ij>\sigma,\alpha}f_{i\sigma,\alpha}^{\dagger}m_{j\sigma,\alpha} 
-t_{B'B'}\sum_{<ij>\sigma,\alpha}m_{i\sigma,\alpha}^{\dagger}m_{j\sigma,\alpha} \\
& & -t_{CC}\sum_{<ij>\sigma,\alpha}f_{i\sigma,\alpha}^{\dagger}f_{j\sigma,\alpha} 
+ J\sum_{i\in Cr} {\bf S}_{i} \cdot
f_{i\alpha}^{\dagger}\vec{\sigma}_{\alpha\beta}f_{i\beta} \\
& & + J_{2}\sum_{i\in Cr,j\in B} {\bf S}_{i} \cdot {\bf s}_{j}
\label{fullham}
\end{eqnarray*}
where the $f$'s and $m$'s refer to the Cr $t_{2g}$ and B$'$ $t_{2g}$ degrees of freedoms. 
$t_{CB'}$, $t_{B'B'}$, $t_{CC}$ represent the nearest neighbor Cr-B$'$, second
nearest neighbor B$'$-B$'$ and Cr-Cr hoppings respectively.  
$\sigma$ is the spin index and $\alpha$ is the orbital index that spans the $t_{2g}$ 
manifold. The difference between the ionic levels,
${\Delta} = \epsilon_{Cr} - \epsilon_{B'}$, defines the on-site energy difference between
Cr $t_{2g}$ and B$'$ $t_{2g}$ levels. $s_{j}$ is the intrinsic moment at the B$^{\prime}$ site. 
The first, six terms of the Hamiltonian, represent the HD mechanism,
which consist of a large core spin at the Cr site ($S_i$) and the coupling between the core spin and
the itinerant electron delocalized over the Cr-B$'$ network.
 Variants of this part has been considered by several authors~\cite{alonso,avignon,Guinea2} in the context of 
SFMO. The last term represents the SE mechanism, that consists of coupling between Cr spin
and the intrinsic moment at B$'$ site. The parameters of the model Hamiltonian
are extracted out of DFT calculations through NMTO downfolding technique of constructing the real space 
Hamiltonian in the basis of effective Cr $t_{2g}$ and B$'$ $t_{2g}$ degrees of freedom.
$t_{CB'}$,  $t_{B'B'}$ and $t_{CC}$ hoppings are found to -0.35 eV, -0.12 eV and -0.08 eV respectively, with
little variation within the W-Re-Os series.
$\Delta$-s show a varying trend within the W-Re-Os series ($\Delta^{W}$ = -0.66 eV, $\Delta^{Re}$ = 0.03 eV, $\Delta^{Os}$ = 0.26 eV). 
The parameters involving $J$ and $J_2$ were obtained from the spin-splitting at Cr site and the extra splitting observed at B$^{'}$ site as compared to that expected from the electron filling effect and the splitting for W compound. 

 The constructed model is then
solved using exact diagonalization on a lattice of dimension 8 $\times$ 8 $\times$ 8. 
Calculations have been carried out as well for lattices of size 4 $\times$ 4 $\times$ 4 and 6 $\times$ 6 $\times$ 6. The 
trend is found to be the same as presented for 8 $\times$ 8 $\times$ 8. 
Exact diagonalization was first carried out
 considering the B$'$ site to be totally nonmagnetic, {\it i.e.} setting the last term to zero, which boils down to the same
 underlying model Hamiltonian as that of SFMO. The energy difference between the PM and FM is found
to decrease with increasing number of valence electron, as shown by diamond symbols in Fig.3. This is exactly the similar trend as found in 
a recent calculation on La doped SFMO,\cite{lsfmo-us,footnote3} as well as in Ref. \onlinecite{millis} with FM getting destabilized with increase of valence electrons. This trend of suppression of $T_c$
upon increasing valence electron count is further amplified
due to change in $\Delta$ within the Cr-B$'$ series. This variation in $\Delta$ in the present series is in contrast to the prescription given 
in ref.\onlinecite{millis} to achieve high $T_c$. Upon reaching valence electron count equal to 3 which corresponds to Os compound, FM becomes totally unstable, reflected in negative sign of the energy difference.  We note a rather rapid decrease in moving from N=2 case to N=3 case. 
This may reflect the special situation of Os compound, with Cr $t_{2g}^{3}$-B$'$ $t_{2g}^{3}$
configuration, an ideal super-exchange situation with insulating solution, that adds on
to the general trend.
The situation gets dramatically changed upon inclusion of the growing localized magnetic nature of the B$'$ site, as shown 
by square symbols in Fig.3.
Considering $J_{2}$ values, as obtained in GGA calculations, we 
find that PM and FM energy difference, recovers the correct trend in moving from W to Re to Os as has been observed experimentally.
Mapping
the PM and FM energy difference to the mean field T$_c$, one obtains values 870 K, 1160 K and 1450 K for the W, Re and Os compounds respectively.
Although the values are overestimated compared to experimental values, presumably due to the finite size effect of exact diagonalization calculation and the mean field formula, the trend is very well reproduced with T$_c^{Re}$/T$_c^{W}$ = 1.33
and T$_c^{Os}$/T$_c^{Re}$ = 1.25, compared to experimental estimates of 1.38 and 1.17 respectively.\cite{tc-w, tc-re, tc-os}
 \begin{figure}
\includegraphics[width=3.4cm]{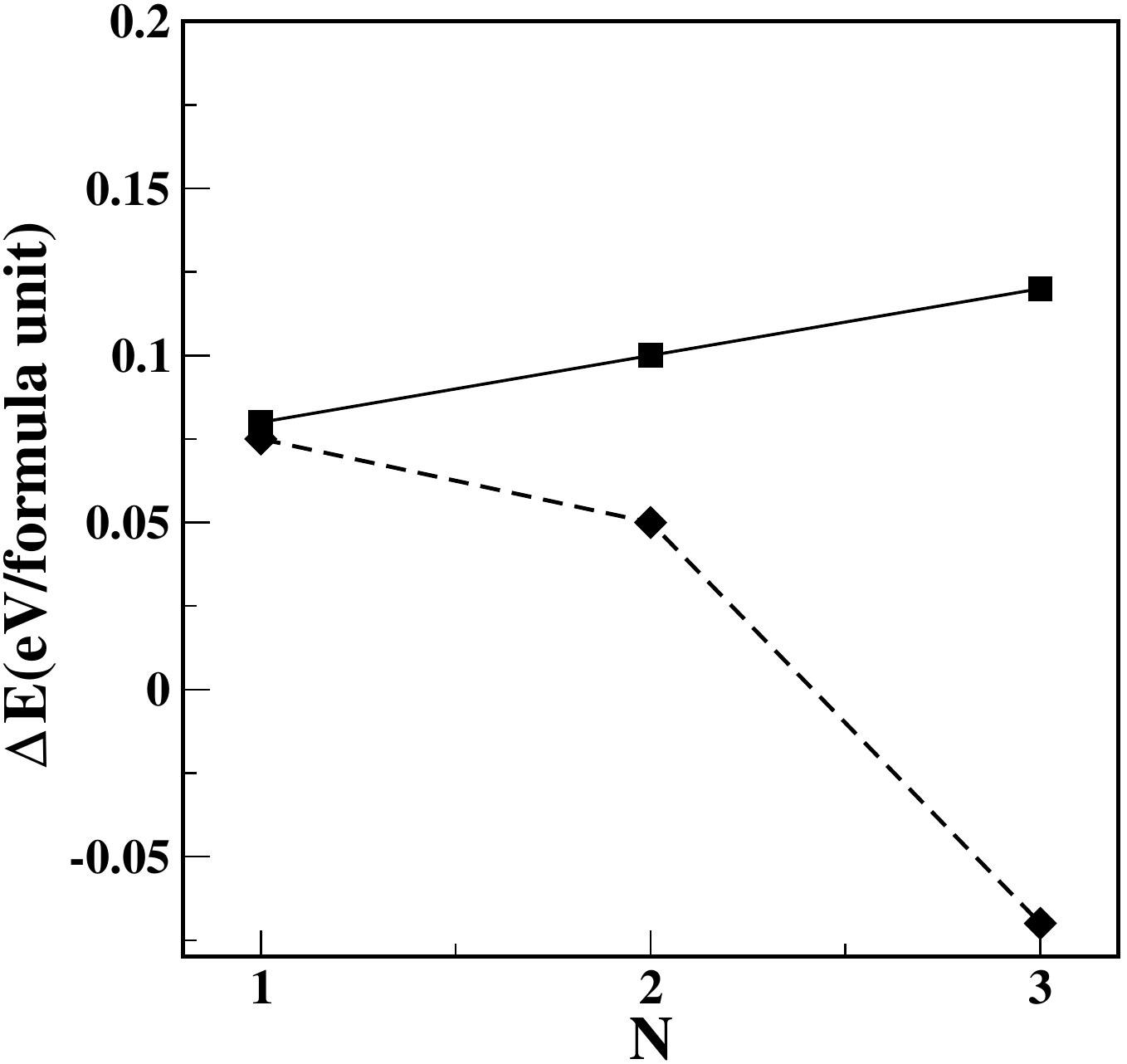}
\caption{\label{fig-3} PM - FM energy differences plotted as a function
of valence electron count, as obtained in exact diagonalization calculation. 
The diamond (square) symbols connected by dashed (solid) line correspond to calculations corresponding to Hamiltonian, without (with) $J_2$ term.}
\end{figure}  

\section{Conclusion and Discussion}

In conclusion, we have studied the counter-intuitive T$_c$ trend in Cr based 
double perovskites, Sr$_2$CrB$'$O$_6$ (B$'$=W/Re/Os).
Analysis of the electronic and magnetic properties shows that the
progressive enhancement of the T$_c$ across the 5d series
should be understood as the interplay of two driving mechanisms: HD mechanism responsible for the
negative spin splitting at B$'$ site as in SFMO,
and SE mechanism. The HD mechanism
gets weaker as one moves along
the series from W to Re to Os, due to the increased energy level separation of B$'$ from Cr. SE, on the other hand, 
gets stronger in moving from SCWO to SCRO to SCOO due to the presence of growing intrinsic moment at B$'$ site, following
the dehybridization effect. The observation of 
uncompensated moment in SCOO arises due to the presence of SO. SCOO, in that sense, should be thought
as a ferrimagnet rather than a ferromagnet.  With this, we demystify the puzzling $T_c$ trend in Cr-B$'$ double perovskite series.

\section{Acknowledgment}
TSD and DDS acknowledge support of DST, Swarnajayanti and J.C.Bose fellowships.
TSD gratefully acknowledges discussion with L. Alff.

$\ast$ tanusri@bose.res.in $\dagger$ sarma.dd@gmail.com


\begin{thebibliography}{99}
\bibitem{footnote3} The experimental observation of increase of $T_c$ upon La doping in SFMO 
may be rationalized in terms of the increase of the antisite disorder as predicted in Ref.\onlinecite{alonso}.
\end{thebibliography}
\end{document}